\documentclass[aps,prb,twocolumn,superscriptaddress,showpacs]{revtex4}

\usepackage{graphicx}
\usepackage{dcolumn}
\usepackage{bm}
\usepackage{amsmath}
\usepackage{amssymb}
\usepackage{latexsym}
\usepackage{epsfig}
\usepackage{amsbsy}
\usepackage{array}
\usepackage{amssymb}
\usepackage{setspace}
\usepackage{bm}

\def\sint{\ifmmode{- \!\!\!\!\!\! \int}
    \else{\hbox{$- \!\!\!\! \int \ $}}\fi}

\begin{document}


\title{First-principles study of the stability of free-standing germanene in oxygen}

\author{G. Liu}

\affiliation{Institute of Laser Engineering, Beijing University of
Technology, Beijing 100124, China}

\affiliation{College of Physics and Communication Electronics,
Jiangxi Normal University, Nanchang 330022, China}

\author{S. B. Liu\footnote{ electronic mail:sbliu@bjut.edu.cn.
}}

\affiliation{Institute of Laser Engineering, Beijing University of
Technology, Beijing 100124, China}

\author{B. Xu}

\affiliation{College of Physics and Communication Electronics,
Jiangxi Normal University, Nanchang 330022, China}

\author{C. Y. Ouyang}

\affiliation{College of Physics and Communication Electronics,
Jiangxi Normal University, Nanchang 330022, China}

\author{H. Y. Song}

\affiliation{Institute of Laser Engineering, Beijing University of
Technology, Beijing 100124, China}

\author{X. L. Li}

\affiliation{Institute of Laser Engineering, Beijing University of
Technology, Beijing 100124, China}


\date{\today}

\begin{abstract}
The O$_{2}$ dissociation and O atoms adsorption on free-standing
germanene are studied by using first-principles calculations in this
letter. Compared with the spontaneous dissociation of oxygen
molecule on free-standing silicene in air, germanene is more stable
than silicene from kinetic point of view, with overcoming energy
barrier of about 0.55 eV. Especially, in contrast with the unique
chemical adsorption of O$_{2}$-dissociation-induced O atoms on
silicene, oxygen molecule can behave a correspondingly stable
adsorption on germanene surface. Moreover, single O atom adsorption
on germanene is also different to that on silicene, resulting in two
opposite migration pathways on germanene surface. Furthermore, once
the oxygen molecule dissociates into O atoms on germanene surface,
the migration and desorption of O atoms are relatively difficult
under room temperature due to the strong Ge-O bonds in the
O-adsorbed germanene, in favor of forming germanium oxides. The
results provide compelling evidence to show that free-standing
gemanene is relatively stable in oxygen,which is different to
silicene essentially.
\end{abstract}

\pacs{73.22.-f, 71.15.Mb}

\pacs{73.22.-f, 71.15.Mb}
\keywords{germanene,germanane, oxygen adsorption, First-principles, stability}

\maketitle

\section{Introduction}  
Graphene, a single layer material of carbon atoms with
two-dimensional (2D) honeycomb structure, has attracted intensive
attention from scientific community to industry field for its
remarkable properties.\cite{1, 2, 3, 4, 5} Since the discovery of
graphene in 2004, other 2D layered materials similar to graphene,
such as silicene and germanene, have gained renewed interest because
it is important for people to look for 2D materials based on
existing silicon or germanium electronic industry. As group-IV
element graphene-like 2D sheet, most of the known features of
germanene resemble those of silicene. Theoretically, density
functional theory (DFT) studies of silicene or germanene have
illustrated that silicon or germanium prefers \emph{sp}$^{3}$
hybridization instead of \emph{sp}$^{2}$, thus, silicene or
germanene is energetically favorable as a low-buckled (LB) structure
with an amplitude of about 0.44 {\AA} and 0.69 {\AA}.\cite{S.
Cahangirov, L. Seixas} Moreover, the band structures in the LB
configurations of silicene and germanene are ambipolar, and their
charge carrier can behave like a massless Dirac fermion at the K
point because of the $\pi$ and $\pi^{*}$ bands that linearly cross
at the Fermi level. It is also found that the electronic and
magnetic properties of silicene and germanene nanoribbons show size
and geometry dependence.\cite{Y. Ding} Furthermore, phonon
calculations  demonstrate the stability of silicene and germanene in
their ground states.\cite{T. Y. Du, Nathanael J. Roome} In addition,
recent studies predict that SOC effect can open a 1.5 meV band gap
in silicene as well as 25 meV in germanene,\cite{C. C. Liu, L.
Seixas} useful to the topological insulators.

Experimentally, using atomic resolved scanning tunneling microscopy
(STM), it has been demonstrated that silicene can be grown on a
close-packed silver surface [Ag (111)] or ZrB$_{2}$ substrate via
direct condensation of a silicon atomic flux onto the single-crystal
substrate in ultrahigh vacuum conditions.\cite{P. Vogt, A.
Fleurence} One year later, it is reported once more on an iridium
(111) surface.\cite{L. Meng} Furthermore, germanane, the fully
hydrogenated germanene, has also been fabricated using a wet
chemistry method in 2013.\cite{E. Bianco} However, germanene, which
has already been predicted to be stable as freestanding novel
monolayer germanium, has remained elusive. It seems to still exist
only in the theory. But quite recently, an exciting experiment by
D\'{a}vila et al \cite{M. E. D¨¢vila} shows that, a two-dimensional
germanium layer, forming several phases, has been grown $\emph{in
situ}$ by dry deposition of germanium onto the Au(111) surface,
similarly to the formation of silicene on Ag(111). One of these
phases displays a clear honeycomb structure with a very weak
corrugation in STM imaging. Additionally, detailed core-level
spectroscopy measurements along with advanced density functional
theory calculations identify this phase as a
$\surd$3$\times$$\surd$3 reconstructed germanene layer on top of a
$\surd$7$\times$$\surd$7 Au(111) surface. Nevertheless, the
resistance to oxidation has not been concerned in their report. To
our best knowledge, the resistance to oxidation of a novel material
is an essential prerequisite for its future applications.

In our previous work,\cite{G. Liu} it shows that free-standing
silicene is extremely active in oxygen due to the spontaneously
dissociation of O$_{2}$ molecule on silicene without overcoming any
energy barrier, implying an exothermic process essentially.
Moreover, the migration and desorption of O atoms are relatively
difficult under room temperature because of the strong Si-O bonds in
the O-adsorbed silicene, which is in favor of forming silicon
oxides. Therefore, in order to take advantage of the good electronic
characters of silicene, some ways to protect the silicene from
exposing to oxygen must be found out, like Molle's\cite{A. Molle}
methodology of encapsulated silicene by oxides-covered
heterostructures. As mentioned above, it naturally raises a
question: Can free-standing genmanene be stable in oxygen?

To answer this question and further understand the details of
interaction between oxygen molecule and germanene surface, the
O$_{2}$ dissociation and O atoms adsorption on free-standing
germanene are studied by using first-principles calculations in this
letter. Our results show that, in comparison to extreme activity of
silicene in oxygen, germanene is relatively stable on account of the
energy barrier of about 0.55 eV to overcome. Moreover, oxygen
molecule can be stably adsorbed on germanene surface, differing in
the case of barely chemical adsorption of O atoms on silicene.
Furthermore, O atoms adsorbed on germanene are relatively difficult
to migrate on or be desorbed from surface, leading to poor mobility
of O atoms.

\section{method}  
All calculations are performed by using the VASP (Vienna ab initio
simulation package) \cite{G. Kresse-1} within the projector
augmented-wave (PAW) approach.\cite{G. Kresse-2} The ground state of
the electronic structure is described within density functional
theory (DFT) using the generalized gradient approximation (GGA) with
Perdew-Burke-Ernzerhof (PBE) exchange correlation
functional.\cite{J. P. Perdew} The energy cutoff for expansion of
wave functions and potentials is 550 eV. The single layer genmanene
is modeled with a 4$\times$4$\times$1 supercell containing 32 Ge
atoms. All of models are separated with a 15 {\AA} vacuum layer in
the z-axis direction. Monkhorst-Pack special k-point method \cite{H.
J. Monkhorst} is used with a grid of 3$\times$3$\times$2. The entire
systems are relaxed by conjugate gradient method until the force on
each atom is less than 0.03 eV/{\AA}. To optimize the O$_{2}$
molecule dissociation path, the climbing image nudged elastic band
(CINEB) method \cite{G. Henkelman} is employed, which has been
proved to be quite effective to calculate the diffusion energy
barrier.

\section{Results and Discussions}

To investigate the stability of germanene in oxygen accurately, the
lattice and geometry of the low-buckled germanene unit cell is
optimized, as shown in Fig .1. The lattice constant a=4.06 {\AA} and
buckling $\Delta$=0.69 {\AA} are obtained. The corresponding DOS
shows the semimetallic or zerogap semiconducting character of
germanene, in agreement with the previous report.\cite{L. Seixas}

To simulate the Ge-O ratio dependance of the O$_{2}$ molecule
adsorption on germanene, one O$_{2}$ molecule is put on unit cells
with different sizes from 1$\times$1$\times$1 to
4$\times$4$\times$1. Results show that for all the unit cell sizes
applied, if the interaction between O$_{2}$ molecule and germanene
is strong enough, O$_{2}$ molecule dissociates into two O atoms that
bond tightly with Ge atoms after relaxation, which indicates that
the process of dissociation is independent of the size of unit cell.
In the following, a 4$\times$4$\times$1 super cell is used to
further investigate the dependence of O$_{2}$ dissociation on
initial positions and orientations of the O$_{2}$ molecule.
Different separation distance between O$_{2}$ and germanene are
tested, which are 0.60, 1.50, 3.00, 4.00 and 6.60 {\AA},
respectively. For each distance, the O$_{2}$ molecule is rotated and
its initial position is also adjusted.

It is found that, when the distance is 6.60 {\AA}, the optimized
adsorption distance decreases to about 6.40 {\AA} and the
orientation of oxygen is barely changed. The O-O bond length
corresponds to 1.25 {\AA}, slightly longer than that of 1.23 {\AA}
in vacuum. The corresponding adsorption energy (defined as the
energy difference between the total energy of O-adsorbed germanene
and the sum of the total energies of the free isolated O$_{2}$
molecule and free-standing germanene) is nearly 0 eV, indicating
that the interaction between the O$_{2}$ molecule and the germanene
is rather weak under this condition. It means a poor physical
adsorption.

While the distance are 4.00 {\AA}, 3.00 {\AA} and 1.50 {\AA},
respectively, results show that oxygen molecule has been adsorbed on
germanene surface after relaxation. The corresponding O-O bond
length, Ge-O bond length and the adsorption distance are about 1.44
{\AA}, 1.94 {\AA} and 1.78 {\AA}, respectively.  The adsorption
energy corresponds to 1.76 eV. Compared with the poor physical
adsorption, it shows that the interaction between oxygen and
germanene surface becomes stronger. On the other hand, it also means
that the O$_{2}$ adsorption is independent of the initial position
and orientation of the molecule for the separation of 1.50 {\AA} and
4.00 {\AA}, implying a spontaneous adsorption of oxygen molecule if
the interaction is strong enough. Importantly, the O-O bond length
of 1.44 {\AA}, the Ge-O bond length of 1.94 {\AA} indicates that the
O-O bond is not broken and Ge atoms bond with O atoms, showing a
chemical adsorption of oxygen molecule on germanene which is so
different to silicene. Especially, the corresponding adsorption
energies of 1.76 eV is quite large for simple physical adsorption.
Therefore, charge transfer might occur when O$_{2}$ molecule is
adsorbed on the germanene surface. In order to verify this, the
induced charge density defined as
$$
\Delta \rho = \rho(Ge_{32}O_{2})-\rho(Ge_{32})-\rho(O_{2})
$$
is plotted in Fig. 2, where $\rho$(Ge$_{32}$O$_{2}$) and
$\rho$(Ge$_{32}$) are the charge density of the germanene system
with and without O$_{2}$ adsorbed on the surface, respectively.
$\rho$(O$_{2}$) is the charge density of O$_{2}$ molecule. With this
definition, isosurfaces of positive values indicate gain charge,
while negative values indicate loss charge during the adsorption
process. It shows clearly from Fig. 2 that O$_{2}$ molecule gains
abundant charge when it is adsorbed on the germanene surface,while
two Ge atoms bonded with two O atoms loss the charge. It indicates
the stronger interaction and bond strength between Ge and O atoms.

As the separation distance decreases to 0.6 {\AA}, the dissociation
of oxygen molecule happens. The oxygen molecule dissociates into two
O atoms on different bridge sites of two nearest neighboring Ge
atoms. The adsorption energy corresponds to 4.08 eV. The
corresponding O-O distance and Ge-O bond length are 3.34 {\AA} and
1.83 {\AA}, respectively, implying that the O-O bond has been
broken.

After dissociation, the most stable configuration of the two O atoms
is calculated to be locating at two bridge sites. To investigate the
detail of O$_{2}$ molecule dissociation on germanene surface and
obtain the optimized dissociation pathway of O$_{2}$, the CINEB
method is applied. As the distance decreases from 4.00 to 3.00 then
to 1.50 {\AA}, the oxygen molecule is all adsorbed on germanene.
Therefore, the optimized adsorption state with the lowest energy
corresponding to separation distance of 3.00 {\AA} can be regarded
as initial state in CINEB method. The O$_{2}$ dissociation state
with the two O atoms locating on two bridge sites can be regarded as
Final state. The state of two O atoms locating on two top site of
germanene is calculated as transition state. To illustrate the
spontaneous adsorption of oxygen on germanene, the evolution from
the free O$_{2}$ molecule to adsorbed oxygen molecule is also
considered in our calculations. The optimized dissociation pathway
of oxygen molecule is shown in Fig. 3. The spontaneous oxygen
molecule adsorption on germanene surface can be seen in Fig.
3(a)-(b), indicating an exothermic process which is similar to the
O$_{2}$ dissociation on silicene. This may originate from gradually
strengthened Ge-O bond, which lowers energy of the system
substantially. But from Fig .3 (b)-(d), the most important
dissociation process of oxygen molecule shows that there is a energy
barrier of about 0.55 eV to overcome. From the thermodynamic point
of view, the O$_{2}$ molecule dissociation reaction on germanene is
also an exothermic process. Kinetically, in comparison with the
spontaneous O$_{2}$ dissociation on silicene, the energy barrier of
about 0.55 eV indicates that freestanding germanene is more stable
than silicene in the presence of O$_{2}$ and its dissociation
reaction is quite different from that of silicene intrinsically. In
this proceeding, Ge-O bond become further strengthened while O-O
bond become further weakened, resulting in the Ge-O and O-O bond
lengths change from 1.94 {\AA}, 1.44 {\AA} to 1.84 {\AA}, 1.89
{\AA}, as shown in Fig. 3 (c). Sequentially, as a result of durative
exothermic process, the dissociation reaction of oxygen molecule
comes from transition state into final state, as shown in Fig. 3
(d). The corresponding O-O distance is also changed from 1.89 {\AA}
to 3.34 {\AA}. Compared with the 1.23 {\AA} in vacuum , the O-O
distance of 3.34 {\AA} shows that the O$_{2}$ totally dissociates on
grenanene. On the other hand, the corresponding adsorption energy of
4.08 eV in final state is lower than that of 5.36 eV on silicene,
demonstrating that the silicon oxide is more stable than germanium
oxide from another perspective.

Since oxygen molecule can dissociate into two O atoms on germanene
surface, the adsorption and the activity of
O$_{2}$-dissociation-induced O atoms are necessary to be discussed
to further understand the stability of the most favorable adsorption
sites. Four typical adsorption sites of the O atom are considered in
our calculations which are the bridge site of two
nearest-neighboring Ge atoms, the top sites of two different Ge
atoms and the center site of Ge$_{6}$ ring, marked with 1, 2, 3, and
4 in Fig. 4(a), respectively. Sites 2 and 3 are different due to the
buckling of the germanene. The adsorption energy ($\emph{E}$$_{ad}$)
is defined as:
$$
E_{ad}=-[E(Ge_{32}O)-E(Ge_{32})-1/2\times E(O_{2})]
$$
where the \emph{E}(\emph{Ge}$_{32}$\emph{O}) and
\emph{E}(\emph{Ge}$_{32}$) are the total energy of the geramnene
supercell model with and without adsorption of one O atom,
respectively.\emph{E}(\emph{O}$_{2}$) is the total energy of O$_{2}$
molecular in vacuum. The calculated adsorption energy
\emph{E}$_{ad}$ and adsorption distance \emph{D }(defined as the
distance from the adsorbate to the substrate plane, namely, the
difference between the \emph{z}-axis coordinate of the adsorbate and
the average of \emph{z}-axis coordinate of all surface atoms) for
the adsorption of an O atom on different sites of germanene are
listed in Table 1.

It is found that the strongest adsorption of one O atom on germanene
is locating at the bridge site (site 1) with adsorption energy of
2.14 eV and adsorption distance of 1.59 {\AA}, corresponding to the
most favorable dissociation sites of the O$_{2}$ molecule, as shown
in Fig. 4 (b). The corresponding Ge-O bond length between the O atom
and the neighboring Ge atom is about 1.84 {\AA}. Interestingly, it
can be seen from Fig. 4 (c) that the O atom adsorbed on top site 2
drops down to but the Ge atom is pulled out of the basin plane of
germanene after relaxation, which is quite different from the case
of top site 3. It may originate from the slightly longer Ge-Ge bond
length and weakened Ge-Ge bond strength, compared with that of
silicene.

To examine the activity of the adsorbed O atoms, the diffusion of
the O atom on germanene is investigated thoroughly. In order to find
the optimized migration pathways, the CINEB method is also applied.
Since the bridge sites are the most favorable adsorption sites and
the migration of the O atom is from one bridge site to the
neighboring bridge site on germanene surface, the corresponding
state of the O atom adsorbed on the bridge site (site 1) can be
regarded as initial state or final state in the CINEB method. In
contrast, the states corresponding to the top sites (site 2 and site
3) can be regarded as different transition states because site 2 is
not equivalent to site 3 in the low-buckled structure. Thus, there
are two migration pathways considered in our calculations, which are
denoted by "1-2-1" and "1-3-1", respectively, as shown in Fig. 5(a)
and (b). The numbers represent the adsorption sites of O atom given
in Fig. 4. The corresponding energy profiles along the optimized
migration pathways are shown in Fig. 5. The calculated energy
barriers are 1.43 eV and 1.28 eV for the two pathways, which are
higher than that of the migration of the O atom on silicene (1.18 eV
and 1.05 eV),\cite{G. Liu} indicating that the mobility of the O
atom on germanene is lower than that on silicene. In addition,
similar to the migration of the O atom on silicene, the O atom moves
along the edge of the string from the bridge site to the top site
and then to the neighboring bridge site. However, it is worth noting
that the migration pathway denoted by "1-2-1" is different from
another migration pathway denoted by "1-3-1". Due to the adsorption
of O atom on site 2 leading O atom to drop down to the basin plane
of germanene after relaxation, O atom moves from bridge site 1 with
higher position to top site 2 with lower position then to bridge
site 1 with higher position. It can explain the difference of
diagrammatic sketch between two migration pathways.

Furthermore, although the calculated results are obtained under
ground state (0K), but even at room temperature (300K) the energy
barrier of about 1.3 eV is not easy to overcome, indicating that the
desorption of the adsorbed O atoms on germanene is relatively
difficult. Therefore, although free-standing gemanene is more stable
than silicene in oxygen, but once oxygen molecule dissociates on
germanene, the O$_{2}$-dissociation-induced O atoms adsorption
prefers to form the germanium oxides.

\section{Conclusion}  
In conclusion, the O$_{2}$ dissociation and O atoms adsorption on
germanene are studied by first-principles calculations in this
letter. Compared with the spontaneous dissociation reaction of
O$_{2}$ on silicene in oxygen, it is found that germanene is more
stable than silicene because oxygen molecule can not easily
dissociate into two O atoms, with overcoming the energy barrier of
about 0.55 eV. On the other hand, oxygen molecule can be adsorbed on
germanene surface with chemical adsorption, although the adsorption
reaction is spontaneous. Moreover, O atom adsorption on germanene is
rather different from the case on silicene, leading to two opposite
migration pathways of O atom on germanene surface. Furthermore, once
the oxygen molecule dissociates into two O atoms on germanene
surface, the migration and desorption of O atoms are relatively
difficult under room temperature due to the strong Ge-O bonds in the
O-adsorbed germanene,which is in favor of forming germanium oxides
in oxygen. The results provide compelling evidence to show that
free-standing gemanene is relatively stable in oxygen,which is
different to silicene substantially. The work is helpful to reveal
the detail of the interaction between oxygen molecule and
free-standing germanene surface, and thus helpful to understand the
stability of germanene in oxygen.

The supports of NSFC under Grant Nos. 11064004, 11234013 and
11264014 are thanked. C. Y. Ouyang is also supported by the "Gan-po
talent 555" Project of Jiangxi Province and the oversea returned
project from the Ministry of Education.


\newpage

\begin{table}[h]
\caption{Calculated adsorption energy \emph{E}\emph{$_{ad}$} and
adsorption distance \emph{D} for the adsorption of an O atom on
different sites of germanene.}
\begin{center}
\begin{tabular}{llllllllllll} 

\hline\hline Sites & 1(bridge)& 2(top1)& 3(top2) &(4(center) \\

\hline \emph{D}({\AA})\ \ & 1.59 & 0.36 & 2.11 & 1.66 \\

\hline \\\emph{E}$_{ad}$(eV)\ \ & 2.14 & 0.84 & 0.69 & 2.13 \\

\hline \hline
\end{tabular}
\end{center} \label{tab:phase}
\end{table}

\newpage







\newpage

\begin{figure}
\centering
\begin{minipage}[b]{0.5\textwidth}
\centering
\includegraphics[width=4in]{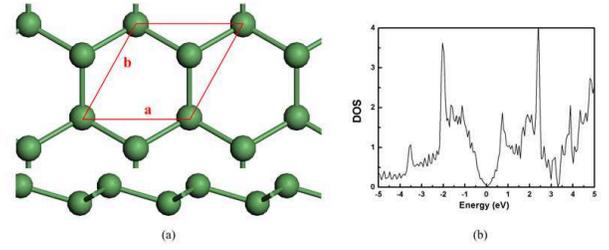}
\end{minipage}
\caption{(Color online) (a) Diagrammatic sketch of top and side
views of germanene, (b) The DOS of unit germanene.
\label{fig:Graph1}}
\end{figure}

\begin{figure}
\centering
\begin{minipage}[b]{0.5\textwidth}
\centering
\includegraphics[width=4in]{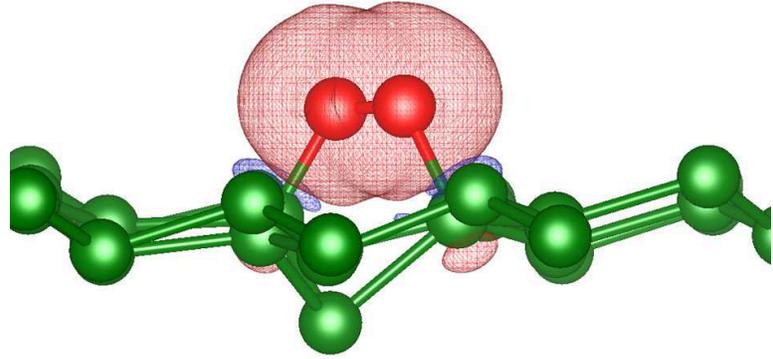}
\end{minipage}
\caption{(Color online) Isosurfaces of the induced charge density
$\Delta$$\rho$ of the germanene adsorbed with an O$_{2}$ molecule.
The green and  red spheres are Ge and O atoms, respectively. The
positive and negative isosurfaces are in rosiness and blue,
respectively. \label{fig:Graph2}}
\end{figure}

\begin{figure}
\centering
\begin{minipage}[b]{0.5\textwidth}
\centering
\includegraphics[width=5in]{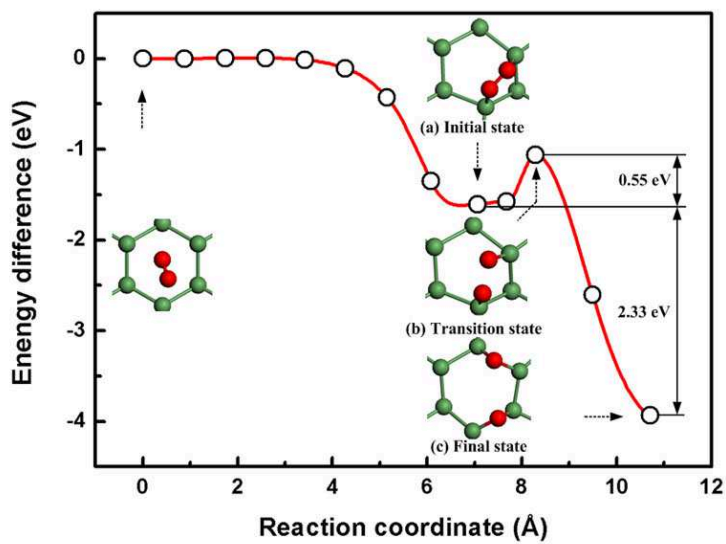}
\end{minipage}
\caption{(Color online) The optimized dissociation pathway of oxygen
molecule on germanene\label{fig:Graph2}}
\end{figure}

\begin{figure}
\centering
\begin{minipage}[b]{0.5\textwidth}
\centering
\includegraphics[width=4in]{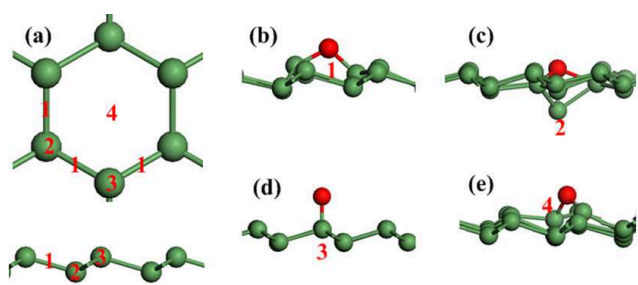}
\end{minipage}
\caption{(Color online)  (a) Diagrammatic sketch of different
adsorption sites of O atom on germanene surface, (b)-(d) represent
optimized adsorption of O atom on bridge site, two different top
sites and center site, respectively. \label{fig:Graph3}}
\end{figure}

\begin{figure}
\centering
\begin{minipage}[b]{0.5\textwidth}
\centering
\includegraphics[width=4in]{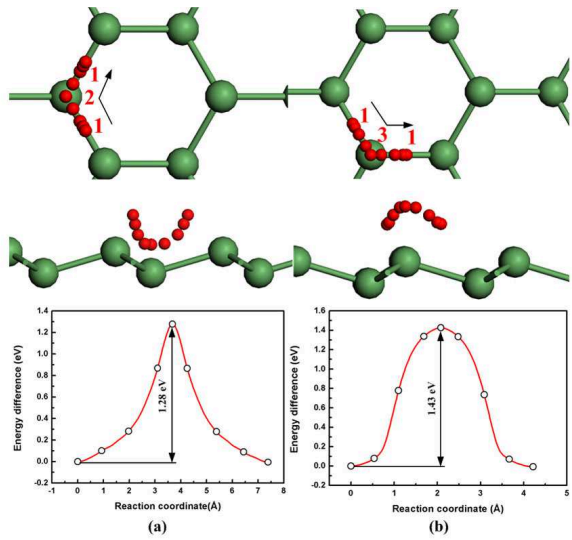}
\end{minipage}
\caption{(Color online) (a) Top and side views of the optimized
migration pathway of an O atom migration along the pathway from site
No. 1 to site No. 2 then to site No. 1 and its energy profile along
the pathway on the germanene surface. (b) Top and side views of the
optimized migration pathway of an O atom migration along the pathway
from site No. 1 to site No. 3 then to site No. 1 and its energy
profile along the pathway on the germanene surface. (Red color and
green color represent O atoms and Ge atoms,
respectively)\label{fig:Graph3}}
\end{figure}

\end{document}